\def\etal{{\it et al.\thinspace}}
\def\eg{{\it e.g.\ }}

\def\ie{{\it i.e.\ }}
\def\cf{{\it c.f.\ }}
\def\gsim{~\rlap{$>$}{\lower 1.0ex\hbox{$\sim$}}}
\def\lsim{~\rlap{$<$}{\lower 1.0ex\hbox{$\sim$}}}

\documentclass[12pt,preprint]{aastex}

\begin{document}

\title{Stability Limits in Resonant Planetary Systems}

\author{Rory Barnes\altaffilmark{1} and Richard Greenberg\altaffilmark{1}}

\altaffiltext{1}{Lunar and Planetary Laboratory, University of Arizona,
Tucson, AZ 85721}


\begin{abstract}
The relationship between the boundaries for Hill and Lagrange
stability in orbital element space is modified in the case of
resonantly interacting planets. Hill stability requires the ordering
of the planets to remain constant while Lagrange stability also
requires all planets to remain bound to the central star. The Hill
stability boundary is defined analytically, but no equations exist to
define the Lagrange boundary, so we perform numerical experiments to
estimate the location of this boundary. To explore the effect of
resonances, we consider orbital element space near the conditions in
the HD 82943 and 55 Cnc systems. Previous studies have shown that, for
non-resonant systems, the two stability boundaries are nearly
coincident.  However the Hill stability formula are not applicable to
resonant systems, and our investigation shows how the two boundaries
diverge in the presence of a mean-motion resonance, while confirming
that the Hill and Lagrange boundaries are similar otherwise. In
resonance the region of stability is larger than the domain defined by
the analytic formula for Hill stability. We find that nearly all
known resonant interactions currently lie in this extra stable region,
\ie where the orbits would be unstable according to the non-resonant
Hill stability formula. This result bears on the dynamical packing of
planetary systems, showing how quantifying planetary systems'
dynamical interactions (such as proximity to the Hill-stability
boundary) provides new constraints on planet formation models.
\end{abstract}

\section{Introduction}
By the end of 2006, 20 multiple planetary systems had been detected
beyond the Solar System (Butler \etal 2006, Wright \etal 2007). Of
these, 7 are likely to contain at least 1 pair that is in a
mean-motion resonance. Barnes \& Quinn (2004; hereafter BQ) showed
that one of these resonant pairs, HD 82943 b and c, had best-fit
orbital elements that placed the system near a stability
limit. Indeed, the best fit was unstable, but a small change (well
within observational uncertainties) in the eccentricity $e$ of the
outer planet would make the system stable (BQ; Ferraz-Mello \etal
2005). BQ also showed that stability requires the ratio of the orbital
periods, $P_c/P_b$, be near 2, and that the relative mean longitudes
and difference in longitudes of pericenter lie in a range such that
conjunctions never occur at the minimum distance between the
orbits. This result suggests that, given the values of $e$ and $a$ of
the two planets, stability is only possible if the two planets are in
the 2:1 resonance.

Two types of stability have been considered in the literature. Hill
stability requires the ordering of planets to remain constant for all
time; the outer planet may escape to infinity. The equations that
define the limits of Hill stability (\ie Marchal \& Bozis 1982;
Gladman 1993) only apply to systems of 2 planets that are not in a
resonance. Lagrange stability requires all planets remain bound to the
star, and the orbits evolve at least quasi-periodically. Lagrange
stability is more meaningful, but its criteria have not been delineated
analytically.

Barnes \& Greenberg (2006a; hereafter BG) showed that the
Hill-stability boundary is nearly the same as the Lagrange-stability
boundary, at least for the non-resonant planets in HD 12661 and 47
UMa. Although the Hill stability boundary was derived for non-resonant
systems, it is not clear how mean-motion resonances distort it. Here
we explore the stability boundary near two resonant systems, HD 82943
(Mayor \etal 2004) and 55 Cnc (Marcy \etal 2002; McArthur \etal 2004). Note that for both
systems the inner planet of the resonant pair is named b and the outer c. We find that the
resonances do provide extra regions of Lagrange stability in phase
space that extend beyond the analytic criterion. In $\S$ 2 we describe
Hill and Lagrange stability and our numerical methods. In $\S$ 3 we
present our results for HD 82943 and 55 Cnc. We also tabulate
proximities to the Hill boundary for all applicable systems and find
that all but one resonantly interacting pair would lie in an unstable
region if not for the resonance. In $\S$ 4 we draw conclusions and
suggest directions for future work.

\section{Methodology}

\subsection{Stability Boundaries}

Hill stability in a coplanar system can be described by the following
inequality:
\begin{equation}
\label{eq:exact}
-\frac{2M}{G^2M_*^3}c^2h > 1 + 3^{4/3}\frac{m_1m_2}{m_3^{2/3}(m_1+m_2)^{4/3}} -
 \frac{m_1m_2(11m_1 + 7m_2)}{3m_3(m_1+m_2)^2} + ...,
\end{equation}
where $M$ is the total mass of the system, $m_1$ is the mass of the
more massive planet, $m_2$ is the mass of the less massive planet,
$m_3$ is the mass of the star, $G$ is the gravitational constant, $M_*
= m_1m_2 + m_1m_3 + m_2m_3$, $c$ is the total angular momentum of the
system, and $h$ is the energy (Marchal \& Bozis 1982). If a given
three-body system satisfies the inequality in Eq.\ (\ref{eq:exact}),
then the system is Hill stable. If this inequality is not satisfied, then the
system may or may not
be Hill stable. In this inequality, the left-hand side is a function of the orbits, but the right-hand side is only
a function of the masses. This approach is fundamentally different
from other common techniques for determining stability which exploit
resonance overlaps (Wisdom 1982; Quillen \& Faber 2006), chaotic
diffusion (Laskar 1990; Pepe \etal 2007), fast Lyapunov indicators
(Froeschl\'e \etal 1997; S\'andor \etal 2007), or periodic orbits
(Voyatzis \& Hadjidemetriou 2005, 2006; Hadjidemetriou 2006).

BG use $\beta$ (the left-hand side of Eq.\ [\ref{eq:exact}]), and
$\beta_{crit}$ (the right-hand side) to describe the Hill stability
boundary.  The Hill stability boundary is the curve defined by
$\beta/\beta_{crit} = 1$. BG showed that the Lagrange stability
boundary appears to be located where $\beta/\beta_{crit}$ is slightly larger than
1 (1.02 for 47 UMa and 1.1 for HD 12661). 

\subsection{Numerical Methods}
For each system, HD 82943 and 55 Cnc, 1000 initial configurations were
generated based on the observational data for each system (Mayor \etal
2004; Marcy \etal 2002), that is, the initial conditions spanned the
range of observational uncertainty.  Note that more recent, improved
elements are available (Butler \etal 2006), but for our purposes the
older values serve equally well. Orbital parameters that have known
errors, such as $e$ and the period, $P$, are varied as a Gaussian
centered on the best fit value, with a standard deviation equal to the
published uncertainty, and orbital elements are sampled appropriately. For
elements with systematic errors, such as inclination, the initial
conditions were varied uniformly. The inclination was varied between 0
and $5^o$, and the longitude of ascending node was varied between 0
and 2$\pi$. Masses were then set to the observed mass divided by the
sine of the inclination. The variation of orbits out of the
fundamental plane will not significantly affect our calculations of
Hill stability (Veras \& Armitage 2004). Each element was varied
independently. The distribution of initial conditions is presented in
Table 1. In this table, $\varpi$ is the longitude of periastron and
$T_{peri}$ is the time of periastron passage. The integrations were
performed with SWIFT (Levison \& Duncan 1994) and MERCURY (Chambers
1999), and conserve energy to at least 1 part in $10^4$. For more
details on these methods, consult BQ.

For each simulation we
numerically determine Lagrange stability on $\sim 10^6$ year
timescales. BQ showed that this timescale identifies nearly all
unstable configurations. We then calculate $\beta/\beta_{crit}$ in
the parameter space sampled by the numerical integrations. Comparison of these two sets of results shows how the Hill and
Lagrange stability boundaries are related near a mean-motion resonance.

\section{Results}
\begin{figure}
\plotone{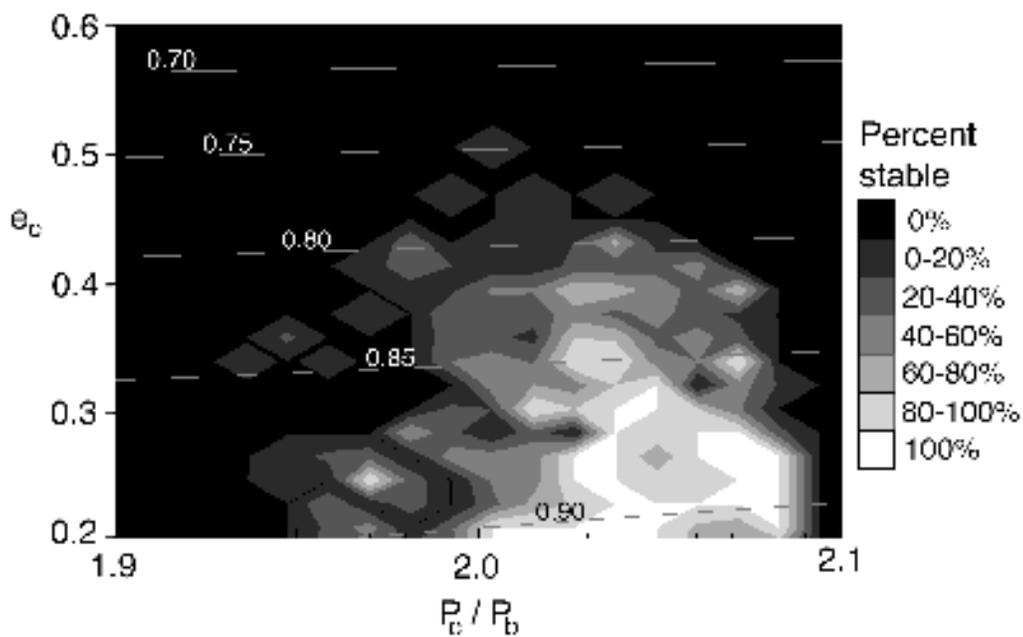}
\figcaption[]{\label{fig:hd82943} \small{Stability map for HD 82943. The shading represents the fraction of simulations that are Lagrange stable on $10^6$ year timescales. The bin sizes are 0.02 for $e_c$ and 0.01 for $P_c/P_b$. Contour lines show values of $\beta/\beta_{crit}$. Ordinarily $\beta/\beta_{crit} < 1.02$ would imply instability, however the mean motion resonance provides a stable region that would not exist if the resonance did not affect the motion.}}
\end{figure}

For HD 82943 the ``stability map'' is shown by the
grayscale shading in Fig.\ \ref{fig:hd82943}.  Shading
indicates the fraction of initial conditions, in a certain range of orbital element space, that give Lagrange stable behavior (no ejections or exchanges) over $10^6$ years:
White bins contained only stable configurations, black only unstable, and
the darkening shades of gray correspond to decreasing fractions of
stable configurations. This representation
plots the stable fraction as a function of two parameters: the eccentricity of the outer planer, $e_c$ and the ratio of the periods $P_c/P_b$. The numerical simulations show that Lagrange stability
is most likely for values of $P_c/P_b$ slightly greater than 2, and
$e_c$ less than 0.4. BQ called this feature the ``stability
peninsula''.

Superimposed on this grayscale map are contours of $\beta/\beta_{crit}$
values.  All the values of $\beta/\beta_{crit}$ are well less than
1.02 which ordinarily would imply instability. However, in the
resonance zone where $P_c/P_b \approx 2$, the stability peninsula
sticks into a regime ($\beta/\beta_{crit}$ as small as 0.75) that would
be unstable if the planets were not in a mean motion resonance. Note
that the numerical simulations include cases with variations of a few
per cent in mass; the $\beta/\beta_{crit}$ contours shown are for the
average mass, but would shift only slightly over the range of masses.

\begin{figure}
\plotone{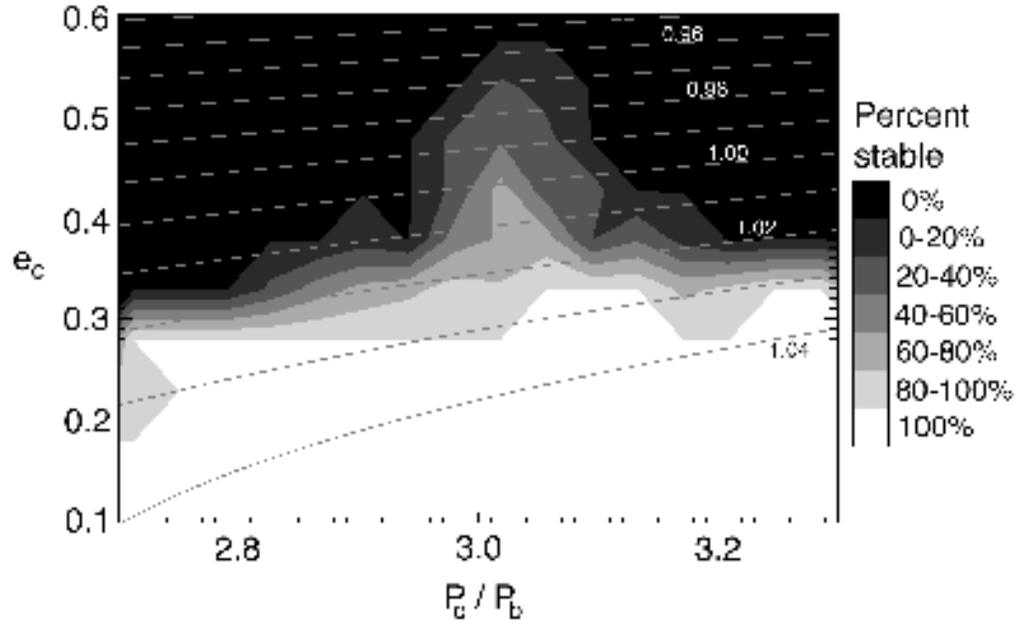}
\figcaption[]{\label{fig:55cnc} \small{Stability map for 55 Cnc (\cf Fig.\ \ref{fig:hd82943}). The bin sizes are 0.05 for $e_c$ and 0.04 for $P_c/P_b$. As in HD 82943, the mean-motion resonance provides a larger Lagrange-stable region.}}
\end{figure}

For 55 Cnc, the stability map (Fig.\ \ref{fig:55cnc}) was developed from integrations over 4 million years, \ie $10^6$
orbits of the outer planet. Our simulations include planets b, c and
d, but not the inner planet, e. Planet e is relatively small and Zhou
\etal (2004) found that the outer, non-resonant planet d does not
appear to affect the global stability of the system. Therefore our
simulations should elucidate the relationship between Hill and
Lagrange stability boundaries in the presence of a 3:1 mean motion
resonance.

\clearpage
\clearpage

\clearpage
\clearpage

Of our simulations $50.2 \pm 5.5$\% were Lagrange stable. The least
massive planet, c, was the planet most likely to be ejected. In this
system we see that stability is likely for $e_c < 0.3$ everywhere,
except at $P_c/P_b \approx 3$, where it extends to $e\sim 0.55$.

Comparing this distribution with the analytical stability criterion
($\beta/\beta_{crit} \lsim 1.03$) we see that the numerical
experiments reproduce that boundary, except in resonance where
$\beta/\beta_{crit}$ can be as low as 0.96. This stability peninsula
for 55 Cnc does not protrude as far into the Hill unstable region as
HD 82943. This difference may be because the 2:1 resonance is of a lower
order (and thus stronger) than the 3:1, and therefore has a more pronounced stabilizing effect. 

Next we tabulate $\beta/\beta_{crit}$ values for all observed systems that
contain two planets. We also include GJ 876 c and b, and $\upsilon$
And c and d. Eq.\ (\ref{eq:exact}) is only applicable to two-planet
systems, but we consider these latter two pairs, which are each part
of a bigger system, as the third planet in each system is probably too small or too
far away to significantly change the interaction of those
pairs. However, interpreting values of $\beta/\beta_{crit}$ in systems
of more than two companions should be made with caution, as there is
no guarantee $\beta/\beta_{crit} = 1$ corresponds to the Hill boundary
for any individual pair.

Table 2 lists values of $\beta/\beta_{crit}$ and the ``class'' of
the interaction, which distinguishes the dominant phenomenon that
changes the shapes of the orbits. ``R'' denotes pairs whose
interaction is dominated by mean-motion resonances (Table 2 also lists
the resonance), ``T'' indicates pairs that may have experienced
significant tidal evolution, and ``S'' indicates pairs with strong
secular interactions (Barnes \& Greenberg 2006b).  All but one
resonantly interacting pair have $\beta/\beta_{crit}$ values less than
1. If not for the resonance, these systems would be unstable.

\begin{figure}
\plotone{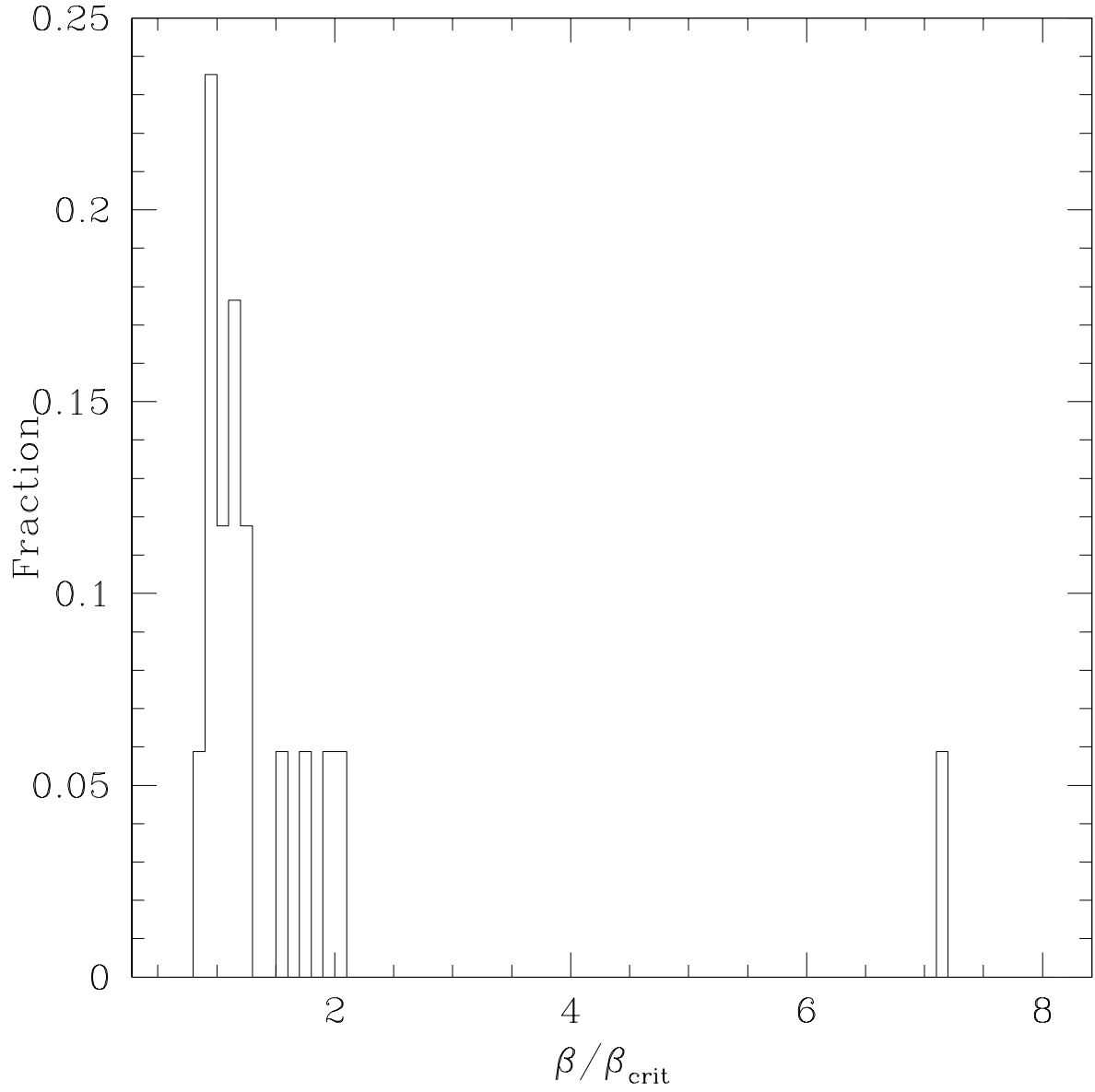}
\figcaption[]{\label{fig:betacrit} \small{Histogram of values of $\beta/\beta_{crit}$ (with a bin size of 0.1) for 17 pairs of planets in Table 2.}}
\end{figure}

Overall, we find 70\% of the pairs we consider are observed to have
$\beta/\beta_{crit} < 1.3$. HD 217107 is observed to have a
$\beta/\beta_{crit}$ significantly larger than other pairs. In Fig.\
\ref{fig:betacrit} we plot the distribution of $\beta/\beta_{crit}$
values from Table 2.

\clearpage
\clearpage

\section{Conclusions}
By explicitly mapping how mean-motion resonances can provide additional
regions of stability in orbital element space, we have found that nearly all observed resonant systems lie in
these extended regions. More generally, we have also shown the distribution of
$\beta/\beta_{crit}$ appears to show that many
planetary systems (resonant or not) lie close to the limits of dynamical
stability. These distributions provide new constraints for models of planet
formation.

In the cases presented here, the 2:1 resonance provides a larger
stable region than 3:1, presumably because it is a lower order
(stronger) resonance. However, for the 5:1 mean-motion resonance in HD
202206, $\beta/\beta_{crit}$ can reach 0.88 and still be stable. So
why is the range of stability for the 3:1 resonance in 55 Cnc so
small? Perhaps if $e_c$ in the 55 Cnc system has values in excess of
0.5, it does interact with the third planet, destabilizing the
system. Future work should investigate the minimum
$\beta/\beta_{crit}$ that allows stability for each resonance. Future
work may also determine if the peninsula we find in 55 Cnc is
truncated due to interactions with 55 Cnc d.

We seek to identify the origin of the shape of the stability
peninsulas in resonant systems. Ideally a general expression will
eventually be developed that describes the Lagrange stability boundary
that applies to planets both in and out of resonance. One avenue of
research is to focus on close approach distances. In the limit of zero
eccentricity, orbits are unstable if they are separated by less than
3.5 mutual Hill radii (Gladman 1993). Therefore we speculate that
systems with approaches within this distance are unstable. For
secularly evolving systems, the orbits change with time and eventually
the planets will line up at the minimum distance between the orbits
over a secular period. Resonances can prevent planets from lining up
at this danger zone, hence the stability peninsulas. This likely
explanation for the shape of the Lagrange boundaries might be a
fruitful direction of future research into planetary system stability.

The distribution of $\beta/\beta_{crit}$ shows that, regardless of the
the presence of mean-motion resonance, many systems have values that
are close to the stability boundary. This trend appears to support the
hypothesis that planetary systems are dynamically ``packed'', \ie that
additional planets could not exist in orbits between those that are
known without destabilizing the system (Barnes \& Quinn 2001; BQ,
Barnes \& Raymond 2004; Raymond \& Barnes 2005; Raymond \etal
2006). Perhaps there is a minimum value of $\beta/\beta_{crit}$ that
would permit the insertion of an additional planet that leaves the
system still stable. In other words it will be interesting to
determine, for a given value of $\beta/\beta_{crit}$, the largest mass
object that could orbit between two planets and still leave the system
stable. Such a relation could produce an analytic criterion for
dynamical packing, which can currently only be estimated numerically
(\eg Menou \& Tabachnik 2003; Dvorak \etal 2003; Rivera \&
Haghighipour 2007).

Past work provides illumination on the possibility that some
minimum value of $\beta/\beta_{crit}$ may define the limit for which
additional planets could be placed between the observed planets. The
HD 168443 system ($\beta/\beta_{crit} = 1.94$) has been shown to be
unable to support even infinitesimal test masses (Barnes \& Raymond
2004). The region between the known planets of HD 169830
($\beta/\beta_{crit} = 1.28$) is chaotic and a planet in that region
is most likely unstable (\'Erdi \etal 2004). On the other hand, HD
38529 ($\beta/\beta_{crit} = 2.07$) could support a Saturn-mass
companion between the known planets (Barnes \& Raymond 2004). These
results suggest $\beta/\beta_{crit} = 2$ may be the critical value.

The packing of the two planets in HD 190360 demands closer
inspection. Although the orbits are more separated and less eccentric
than those in HD 168443, their $\beta/\beta_{crit}$ value (1.70) is
less than that for HD 168443 (1.94). To explore this issue, we have
integrated the HD 190360 system with a hypothetical Earth-mass planet
on a circular orbit located at the midpoint between the apoastron
distance of the inner planet and the periastron distance of the
outer. The additional companion in the HD 190360 system survived for
$10^6$ years. A similar experiment with HD 74156 ($\beta/\beta_{crit}
= 1.54$) showed ejection of the Earth-mass planet in only 2500
years. We tentatively conclude that systems are packed if
$\beta/\beta_{crit} \lsim 1.5$, not packed if $\beta/\beta_{crit}
\gsim 2$, and the packing status is unknown in the range $1.5 \lsim
\beta/\beta_{crit} \lsim 2$. Future research needs to determine the
relationship between $\beta/\beta_{crit}$ and the possibility that
additional planets could be stable between known ones.

As noted at the end of $\S$ 3.3, 70\% of the tabulated systems have
$\beta/\beta_{crit} < 1.3$, indicating that the planets are too fully
packed to allow any intermediate planets. This result, coupled with
the limitations of radial velocity surveys to detect planets (\eg we
used minimum masses), suggests that the vast majority of multiple
planet systems are similarly fully packed. Our results are therefore
consistent with the ``Packed Planetary Systems'' hypothesis (BQ;
Barnes \& Raymond 2004; Raymond \& Barnes 2005; Raymond \etal 2006;
see also Laskar 1996) which proposes that planetary systems tend to
form so as to be dynamically packed. This hypothesis therefore
predicts that HD 190360 and especially HD 217107 harbor additional,
undetected planets.

This investigation has identified a simple way to parameterize
multiple planet systems. At least for a two-planet system, a single
parameter $\beta/\beta_{crit}$ may indicate both stability and
packing. Moreover, the statistics of the distribution of this
dynamical parameter for observed systems are intriguing: Planetary
systems tend to be dynamically fully packed and resonant systems lie
at values of $\beta/\beta_{crit}$ that would indicate instability for
non-resonant systems. Describing planetary systems by parameterizing
the character of their dynamical interaction is also the approach
taken by Barnes \& Greenberg (2006b), who calculated the proximities
of planetary systems to the apsidal separatrix. These new
methodologies focus on the proximities of the dynamical interactions
to boundaries between qualitatively different dynamical regimes.

It now appears that about half of stars with planets have multiple
planets (Wright \etal 2007), and descriptions of their dynamical
interactions will therefore become increasingly more relevant,
especially since many planets' eccentricities oscillate by two orders
of magnitude (Barnes \& Greenberg 2006b). We encourage research that
models planet formation (\eg Lee \& Peale 2002; S\'andor \& Kley 2006)
to include comparisons of the simulated values of $\beta/\beta_{crit}$
to those of real planetary systems.

\medskip
J. Bryan Henderson, Thomas R. Quinn, and Chance Reschke assisted with
the simulations presented here. An anonymous referee provided helpful
suggestions. This work was funded by NASA's PG\&G program.

\references
Barnes, R. \& Greenberg, R. 2006a, ApJ, 647, L163\\
------------. 2006b, ApJ, 652, L53\\
Barnes, R. \& Quinn, T.R. 2001, ApJ, 554, 884\\ 
------------. 2004, ApJ, 611, 494\\
Barnes, R. \& Raymond, S.N. 2004 ApJ, 617, 569\\
Butler, R.P. \etal 2006, ApJ, 646, 505\\
Chambers, J., 1999, MNRAS, 304, 793\\
Cochran, W. \etal 2007, ApJ in press\\
Correia, A.C.M. \etal 2005, A\&A, 440, 751\\
Dvorak, R. \etal 2003, A\&A, 410, L13\\
\'Erdi, B. \etal 2004, MNRAS, 351, 1043\\
Ferraz-Mello \etal 2005, ApJ, 621, 473\\
Froeschl\'e, C. \etal 1997, CeMDA, 67, 41\\
Levison, H.F. \& Duncan, M.J. 1994, Icarus, 108, 18\\
Gladman, B. 1993, Icarus, 106, 247\\
Hadjidemetriou, D. 2006, CeMDA, 95, 225\\
Laskar, J. 1990, Icarus, 88, 266\\
---------. 1996, CeMDA, 64, 115\\
Lee, M.-H. \& Peale, S.J. 2002, 567, 596\\
Marchal, C. \& Bozis, G. 1982, CeMDA, 26, 311\\
Marcy, G. \etal 2002, ApJ, 581, 1375\\
Mayor, M. \etal 2004, A\&A, 415, 391\\
McArthur, B.E. 2004, ApJ, 614, L81\\
Menou, K. \& Tabachnik, S. 2003, ApJ, 583, 473\\
Murray, C.D. \& Dermott, S.F. 1999 \textit{Solar System Dynamics}, Cambridge UP, Cambridge\\ 
Pepe, F. \etal 2007, A\&A, 462, 769\\
Quillen, A.C, \& Faber, P. 2006, MNRAS, 373, 1245\\
Raymond, S.N. \& Barnes, R. 2005, ApJ, 619, 549\\
Raymond, S.N., Barnes, R. \& Kaib, N.A. 2006, ApJ, 644, 1223\\
Rivera, E.J. \& Haghighipour, N. 2007, MNRAS, 374, 599\\
Veras, D., \& Armitage, P. 2004, Icarus, 172, 349\\
S\'andor, Zs. \& Kley, W. 2006, A\&A, 451, L31\\
S\'andor, Zs. \etal 2007, MNRAS, 375, 1495\\
Santos, N.C., Israelian, G., \& Mayor, M. 2000, A\&A, 363, 228\\
Wright, J.T. \etal 2007, ApJ, 657, 533\\
Voyatzis, G. \& Hadjidemetriou, D. 2005, CeMDA, 93, 263\\
---------. 2006, CeMDA, CeMDA, 95, 259\\
Zhou, L. \etal 2004, MNRAS, 350, 1495\\

\bigskip
\begin{center}Table 1: Orbital Elements and Errors for HD 82943 and 55 Cnc\\
\begin{tabular}{cccccccc}
\hline
System & $m_3$ (M$_\odot$) & Planet & $m$ (M$_{Jup}$) & $P$ (d) & $e$ & $\varpi$ ($^o$) & $T_{peri}$ (JD)\\
\hline\hline
HD 82943 & $1.05 \pm 0.05^a$ & b & 0.88 & $221.6\pm2.7$ & $0.54\pm0.05$ & $138\pm13$ & $2451630.9\pm5.9$ \\
 & & c & 1.63 & $444.6\pm8.8$ & $0.41\pm0.08$ & $96\pm7$ & $2451620.3\pm12.0$ \\
55 Cnc & $0.95\pm0.1^b$ & b & $0.84\pm0.07$ & 14.653$\pm$0.0006 & 0.02$\pm$0.02 & 99$\pm$35 & 2450001.479$\pm 10^{-6}$\\
 & & c & $0.21\pm0.04$ & 44.276$\pm$0.021 & 0.339$\pm$0.21 & 61$\pm$25 & 2450031.4$\pm$2.5\\
 & & d & $4.05\pm0.9$ & 5360$\pm$400 & 0.16$\pm$0.06 & 201$\pm$22 & 2785$\pm$250\\

\end{tabular}
\end{center}
$^a$ Santos \etal (2000); $^b$ Marcy \etal (2002)\\

\bigskip
\begin{center}Table 2: Values of $\beta/\beta_{crit}$ for Known Systems\\
\begin{tabular}{cccc}
\hline
System & Pair & $\beta/\beta_{crit}$ & Class\\
\hline\hline
HD 202206$^a$ & b-c & 0.883 & R (5:1)\\
HD 128311$^b$ & b-c & 0.968 & R (2:1)\\
HD 82943$^b$ & b-c & 0.946 & R (2:1)\\
HD 73526$^b$ & b-c & 0.982  & R (2:1)\\
GJ 876$^b$ & c-d & 0.99$^c$ & R (2:1)\\
47 UMa$^b$ & b-c & 1.025 & S\\
HD 155358$^d$ & b-c & 1.043 & S\\
HD 108874$^b$ & b-c & 1.107 & R (4:1)\\
$\upsilon$ And$^b$ & c-d & 1.125$^c$ & S\\
HD 12661$^b$ & b-c & 1.199 & S\\
HIP 14810$^e$ & b-c & 1.202 & T\\
HD 169830$^b$ & b-c & 1.280 & S\\ 
HD 74156$^b$ & b-c &  1.542 & S\\
HD 190360$^b$ & b-c & 1.701 & T\\
HD 168443$^b$ & b-c & 1.939 & S\\
HD 38529$^b$ & b-c & 2.070 & S\\
HD 217107$^b$ & b-c & 7.191 & T\\
\end{tabular}
\end{center}
$^a$ Correia \etal (2005); $^b$ Butler \etal (2006); $^c$ An
additional planet is present in this system; $^d$ Cochran \etal
(2007); $^e$ Wright \etal (2007)
\end{document}